\documentclass[journal]{IEEEtran}

\usepackage[pdftex]{graphicx}

\ifCLASSINFOpdf
\else
\fi

\usepackage{amsmath}

\usepackage{amsthm}
\usepackage{amssymb}

\usepackage{color}
\usepackage{ifthen}
\usepackage[normalem]{ulem}
\usepackage{soul}

\newcommand{\CTLOPERATOR}[1]{\mbox{\textsf{\textup{#1}}}}

\newcommand{\CTLAG}{\CTLOPERATOR{AG}}

\newcommand{\RISCV}{\mbox{RISC-V}}

\definecolor{newtextcolor}{rgb}{0,0,0.5}

\definecolor{oldtextcolor}{rgb}{0.4,0.4,0.4}

\newcommand{\CODEVAR}[1]{\mbox{\textrm{\textit{#1}}}}
\newcommand{\CODEVARSUB}[2]{\mbox{${\CODEVAR{#1}}_{#2}$}}

\newcommand{\DSOC}{\mbox{$d_{\mbox{\scriptsize SOC}}$}}
\newcommand{\DMEM}{\mbox{$d_{\mbox{\scriptsize MEM}}$}}

\newcommand{\PALERT}{\mbox{P-alert}}
\newcommand{\PALERTS}{\mbox{P-alerts}}
\newcommand{\LALERT}{\mbox{L-alert}}
\newcommand{\LALERTS}{\mbox{L-alerts}}

\newcommand{\ASSCODE}[1]{\texttt{\small #1}}
\newcommand{\REGX}[1]{\ASSCODE{x#1}}

\theoremstyle{definition}
\newtheorem{definition}{Definition}

\begin{document}
\title{\huge Processor Hardware Security Vulnerabilities and their
  Detection by Unique Program Execution Checking\vspace{1ex}}

\author{%
\IEEEauthorblockN{%
  Mohammad Rahmani Fadiheh\IEEEauthorrefmark{1},  
  Dominik Stoffel\IEEEauthorrefmark{1},  
  Clark Barrett\IEEEauthorrefmark{3},
  Subhasish Mitra\IEEEauthorrefmark{2}\IEEEauthorrefmark{3},
  Wolfgang Kunz\IEEEauthorrefmark{1}%
}
\IEEEauthorblockA{%
  \hfill
  \begin{minipage}{0.38\linewidth}
    \normalsize
    \vspace{1ex}
    \begin{center}
      \IEEEauthorrefmark{1}Dept. of Electrical and Computer Engineering\\
      Technische Universit\"at Kaiserslautern,\\ Germany
    \end{center}
  \end{minipage}
  \hfill
  \begin{minipage}{0.3\linewidth}
    \normalsize
    \vspace{1ex}
    \begin{center}
      \IEEEauthorrefmark{2}Dept. of Electrical Engineering\\
      Stanford University, Stanford, CA, USA
    \end{center}
  \end{minipage}
  \hfill
  \begin{minipage}{0.3\linewidth}
    \normalsize
    \vspace{1ex}
    \begin{center}
      \IEEEauthorrefmark{3}Dept. of Computer Science\\
      Stanford University, Stanford, CA, USA
    \end{center}
  \end{minipage}
  \hfill\mbox{}
  \vspace{-0.7mm}
}%
}%

\maketitle

\begin{abstract}
  
Recent discovery of security attacks in advanced processors, known as Spectre and Meltdown, has resulted in high public alertness about security of hardware. %
The root cause of these attacks is 
information leakage across ''covert channels'' that reveal secret data without any explicit information flow between the secret and the attacker. %
Many sources believe that such covert channels are intrinsic to highly advanced processor architectures based on speculation and out-of-order execution, 
suggesting that such security risks can be avoided by staying away from high-end processors. %
This paper, however, shows that the problem is of wider scope: we %
present new classes of covert channel attacks which are possible in average-complexity processors with in-order pipelining, %
as they are mainstream in applications ranging from Internet-of-Things to Autonomous Systems. %

We present a new approach as a foundation for remedy against covert channels: while all previous attacks were found by clever thinking of human attackers, 
this paper presents an automated and exhaustive method called ``Unique Program Execution Checking'' which detects and locates vulnerabilities to covert 
channels systematically, including those to covert channels unknown so far. %

\end{abstract}

\IEEEpeerreviewmaketitle

\section{introduction}
\label{sec:introduction}

Subtle behaviors of microprocessors, notable at the level of their
hardware (HW) implementation in digital logic, are the root cause of
security breaches demonstrated in the
Spectre~\cite{2018-KocherGenkin.etal} and
Meltdown~\cite{2018-LippSchwarz.etal} attacks. %
In design flows commonly used in industry, the digital designer
describes the logic behavior of the processor in a clock~cycle-accurate way and defines for each instruction the elementary steps of
its execution based on the processor's registers. %
Such design descriptions at the ``Register-Transfer Level (RTL)'' are
often referred to as the \emph{microarchitecture} of a processor. %
Numerous degrees of freedom exist for the designer in choosing an
appropriate microarchitecture for a specification of the processor
given at the level of its instruction set architecture (ISA). %

However, the same degrees of freedom that the designer uses for
optimizing a processor design may also lead to \emph{microarchitectural side effects}
that can be exploited in security attacks. %
In fact, it is possible that, depending on the data that is processed,
one and the same program may behave slightly differently in terms of
what data is stored in which registers and at which time points. %
These
differences only affect detailed timing at the microarchitectural
level and have no impact on the correct functioning of the program at
the ISA level, as seen by the programmer. %
However, if these subtle
alterations of program execution at the microarchitectural level can
be caused by secret data, this may open a ``side channel''. %
An attacker may trigger and observe these alterations to infer secret
information. %

In microarchitectural side channel attacks, the possible leakage of
secret information is based on some microarchitectural resource which
creates an information channel between different software (SW)
processes that share this resource. %
For example, the cache can be such a shared resource, and various
attacking schemes have been reported which can deduce critical
information from the footprint of an encryption software on the
cache~\cite{2017-YaromGenkin.etal,2005-Percival,2011-GullaschBangerter.etal,2014-YaromFalkner}. %
Also other shared resources can be (mis-)used as channel in
side-channel attacks, as has been shown for
DRAMs~\cite{2015-PesslGruss.etal} and other shared functional
units~\cite{2007-AciicmezSeifert}. %

In these scenarios, the attacker process alone is not capable of
controlling both ends of a side channel. %
In order to steal secret information, it must interact with another
process initiated by the system, the ``victim process'', which
manipulates the secret. %
This condition for an attack actually allows for remedies at the SW
level which are typically applied to security-critical SW components
like encryption algorithms. %
Common measures include constant time
encryption~\cite{2012-JayasingheRagel.etal} and cache access pattern
obfuscation~\cite{2008-KongAciicmez.etal}. %
They prohibit the information flow at one end of the channel which is
owned by the victim process. %

This general picture was extended by the demonstration of the
Spectre~\cite{2018-KocherGenkin.etal} and
Meltdown~\cite{2018-LippSchwarz.etal} attacks. %
They constitute a new class of microarchitectural side channel
attacks which are based on so called ``covert channels''. 
These are special cases of microarchitectural side channels in which
the attacker controls \emph{both} ends of the channel, the part that
triggers the side effect and the part that observes it. %
In this scenario, a single user-level attacker program
 can establish a microarchitectural side
channel that can leak the secret although it is not manipulated by any
other program. %
Such HW covert channels not only can corrupt the
usefulness of encryption and secure authentication schemes, but can
steal data essentially anywhere in the system. %

This paper presents new covert channels in average complexity processors that can have
severe implications for a wide range of applications in Embedded Systems and Internet-of-Things (IoT) where simple in-order processors are commonly used.  
Our results show that HW vulnerabilities by covert channels are not only a
consequence of early architectural decisions on the features of a processor, such as out-of-order execution or speculative execution. %
In fact, vulnerabilities can also be introduced at a later design stage in the
course of microarchitectural optimizations, targeting speed and
energy, for example. %

Clearly, it cannot be expected from a designer to anticipate all
``clever thinking'' of potential attackers who attempt to create
covert channels. %
Therefore, this paper is dedicated to presenting a new technique which
\emph{automatically} detects all microarchitectural side effects and
points the designer to the HW components that may be involved in the
possible creation of a covert channel. %

\begin{figure}
  \centering
  \includegraphics[trim={36mm 82mm 35mm 71mm}, clip, width=1.0\linewidth]{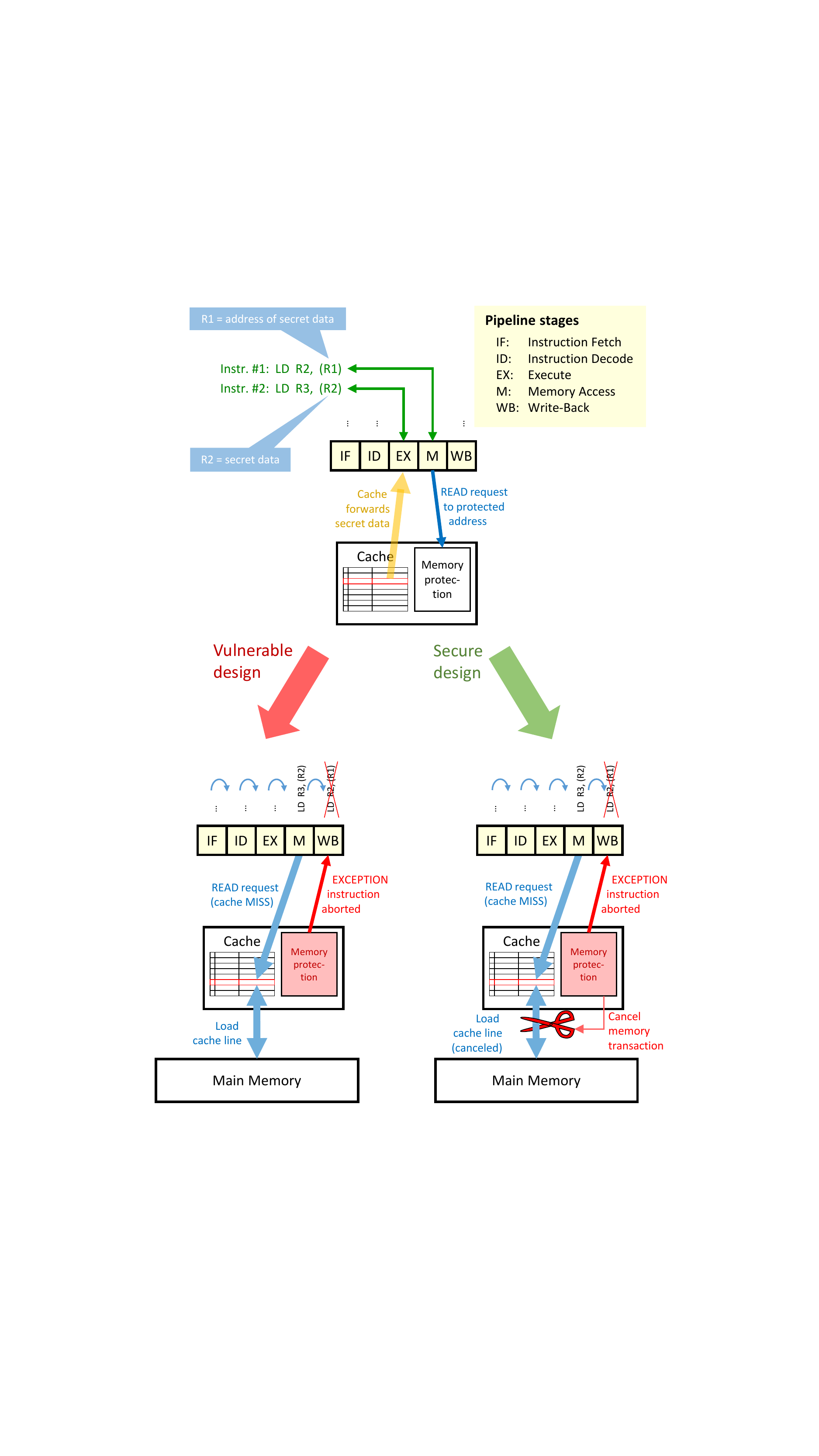}
  \caption{%
    In-order pipeline vulnerability: the design decision whether or
    not a cancellation mechanism is implemented for the transaction
    between memory and cache does neither have an impact on functional
    design correctness nor on the correct implementation of the memory
    protection. %
    However, it may insert a microarchitectural side effect that can
    form the basis for a covert channel-based attack. %
  } %
  \label{fig:in-order-vulnerability}
\end{figure}

Fig.~\ref{fig:in-order-vulnerability} illustrates how a HW
vulnerability may be created by certain design decisions even
in a simple in-order pipeline. %
Consider two instructions as they are executed by a 5-stage pipeline,
and let us assume that register R1 contains the address of some secret
data that should not be accessible for a user-level program. %
(Note that the address itself can be routinely processed by any user
program.) %
Let us consider the routine situation where the cache holds a copy of
the secret data from an earlier execution of operating system-level
code. %
The first instruction, as it enters the memory (M) stage of the
pipeline, makes a READ request to the protected address, attempting to
read the secret data. %
The second instruction at the same time enters the execute (EX) stage
of the pipeline and uses the secret data in an address calculation. %
For performance reasons, a cached copy of the requested secret data may
be forwarded directly to the EX stage. %
This is a common and correct microarchitectural
forwarding~\cite{2013-PattersonHennessy} feature and, by itself, doesn't cause a
problem if the memory protection scheme in place makes sure that
neither instruction~\#1 nor any of the subsequent instructions in the
pipeline can complete. %
We assume that such a memory protection scheme is correctly
implemented in our scenario.%

In the next clock cycle, as shown in the two bottom parts of
Fig.~\ref{fig:in-order-vulnerability}, instruction~\#1 enters the
write-back (WB) stage and instruction~\#2 enters the M stage of
the pipeline. %
Concurrently, the memory protection unit has identified in this clock
cycle that instruction~\#1 attempts access to a protected location. %
This raises an exception for illegal memory access and causes the
pipeline to be flushed. %
From the perspective of the programmer at the ISA level the shown
instructions are ignored by the system. %
In place of these instructions a jump to the exception handler of the
operating system (OS) is executed. %

However, depending on certain design decisions, the instruction
sequence may have a \emph{side effect}. %
Instruction~\#2 triggers a READ request on the cache. %
Let us assume that this request is a \emph{cache miss}, causing the
cache controller to fill the corresponding cache line from main
memory. %
Even though the READ request by the pipeline is never fulfilled
because it is canceled when the exception flushes the pipeline, it may
have measurable consequences in the states of the microarchitecture. %
In the scenario on the left (``vulnerable design'') the state of the
cache is changed because the cache line is updated. %
  
Such a cache foot print dependent on secret data is the basis
for the Meltdown attack. %
It can be exploited as detailed
in~\cite{2018-LippSchwarz.etal}. %
However, in Meltdown this critical step to open a covert channel is
only accomplished exploiting the out-of-order execution of instructions. %
In our example, however, the behavior of the cache-to-memory interface
may cause the security weakness. %
Not only the core-to-cache
transaction but also the cache-to-memory transaction must be canceled
in case of an exception to avoid any observable side effect (``secure
design'' variant on the right). %
This is not at all obvious to a designer. %
There can be various reasons why designers may choose not to
implement cancellation of cache updates. %
First of all, both design variants are functionally correct. %
Moreover, the vulnerable design prohibits access to protected memory
regions. %
Implementing the cancellation feature incurs additional cost in terms
of design and verification efforts, hardware overhead, power
consumption, etc. %
There is no good reason for implementing the feature unless the
designer is aware of the vulnerability. %
  
This example shows that a Meltdown-style attack can be based on even
subtler side effects than those resulting from out-of-order
execution. %
This not only opens new possibilities for known 
Meltdown-style attacks in processors with in-order pipelines. %
These subtle side effects can also form the basis for
new types of covert channel-based attacks which 
have been unknown so far (as demonstrated in Sec.~\ref{sec:raw-hazard-attack}). %

The key contributions of this paper are summarized as follows. %
\begin{itemize}
\item This paper, for the first time, presents the experimental evidence
  that new kinds of covert channel attacks are also possible in
  simple in-order processors. %
  We present the \emph{Orc attack}, which uses a so far
  unknown type of covert channel. %

\item We present a method for security analysis by Unique Program
  Execution Checking (UPEC). %
  UPEC employs a rigorous mathematical (“formal”) analysis on
  the microarchitectural level (RTL). %
  By employing the proposed UPEC methodology the designer can
  precisely assess during design time to what extent hardware security
  is affected by the detailed design decisions. %
  
\item Based on UPEC, for the first time, covert channels can be
  detected by a systematic and largely automated analysis rather than
  only by anticipating the clever thinking of a possible attacker. %
  UPEC can even detect previously unknown HW vulnerabilities, as
  demonstrated by the discovery of the Orc attack in our
  experiments. %

\end{itemize}

\section{Related Work}
\label{sec:related-work}

Information flow tracking (IFT) has been widely used in the field of
security for HW/SW systems. %
Its core idea is to enhance the hardware and/or the software in a way
that the flow of information is explicitly visible. %
Different techniques have been proposed in order to instrument the
processor with additional components enabling it to monitor
information flows and possibly block the illegal flows at run
time~\cite{2004-SuhLee.etal,2009-TiwariWassel.etal}. %
Also CC-hunter~\cite{2014-ChenVenkataramani} targets illegal
information flows at run time. It instruments a processor with a
special module called covert channel auditor which uncovers covert
channel communications between a trojan and a spy process through
detecting specific patterns of conflicting accesses to a shared
resource~\cite{2014-ChenVenkataramani}. %
All these methods incur high overhead on the design and demand
modifications at different levels of the system, such as in the
instruction set architecture (ISA), the operating system (OS) and the
HW implementation. %
CC-hunter also requires defining the conflict indicator event for each
resource which can open a covert channel. This demands a priori
knowledge about possible covert channels. %
In order to capture and remove timing-based side channels in the
design the gate-level netlist can be instrumented with IFT
capabilities~\cite{2014-ObergMeiklejohn.etal}. %
Such a gate-level IFT method is meant to be applied to selected
modules such as a crypto-core or a bus controller. %
It faces complexity problems when addressing security issues of the
larger system. %
Moreover, since the underlying analysis is based on simulation,
relevant corner cases may be missed. %

Ardeshiricham et al.~\cite{2017-ArdeshirichamHu.etal} developed an RTL
transformation framework, called RTLIFT, for enhancing an existing RTL
model with information flow tracking (IFT). %
The method is mainly based on gate-level
IFT~\cite{2009-TiwariWassel.etal} and has been extended by the
Clepsydra approach~\cite{2017b-ArdeshirichamHu.etal} to covering
information flow through timing channels. %
However, any approach based on inserting constant-time operations into
the design, due to the induced overhead, cannot provide a universal
solution to the existence of microarchitectural attacks, such as
Meltdown~\cite{2014-YaromFalkner}. %

\textit{Software security} has been seen as a crucial factor in overall system
security and security specification and verification have long been
studied by the software
community~\cite{2010-OheimbMoedersheim,2004-ZaninMancini}. %
One of the main challenges in this field is the inability of classical
property languages such as CTL to specify the needed security
requirements~\cite{2010-ClarksonSchneider}. %
Hyperproperties and also HyperLTL and HyperCTL* have been proposed for
specifying and verifying security requirements of
software~\cite{2010-ClarksonSchneider, 2014-ClarksonFinkbeiner.etal}. %
Hyperproperties provide the possibility of effectively specifying
general security requirements such as non-interference and
observational determinism~\cite{2010-ClarksonSchneider}. %
Detecting security vulnerabilities and verifying information flow in software are also targeted by the methods of taint analysis and symbolic execution, as surveyed in~\cite{2010-SchwarzAvgerinos.etal}. %

These works have proven to be effective in finding
vulnerabilities in software. %
However, the security threats in HW/SW systems are not
limited to the software alone. %
Vulnerabilities can emerge from HW/SW
interaction or from the hardware itself. %
Also, certain SW vulnerabilities are exploitable through the
existence of microarchitectural features (e.g., cache side channel
attacks). %
This category of security threats may only be visible at a specific
level of hardware abstraction. %
Although the security requirements covering such issues are related
to the ones addressed in SW verification, the SW security
notions cannot be easily adopted in the domain of hardware and
HW-dependent software. %
Open questions such as how to model hardware and the information flow in hardware, how to formulate security requirements for hardware and how to handle the
computational complexity pose significant challenges. Especially after Spectre and Meltdown have created awareness for these questions, this has sparked new and intensive research activities in the domain of hardware security. %

Instruction level abstraction (ILA) is an approach to create a
HW-dependent model of the software and formally verify security
requirements~\cite{2016-SubramanyanMalik.etal}. %
This method has the advantage of capturing vulnerabilities which are
not visible in the software/firmware itself, but rather manifest
themselves through the communication between software and hardware. %
However, HW security vulnerabilities in the RTL implementation
of the hardware cannot be detected by this approach. %

The use of formal techniques for security verification in hardware was
pioneered in~\cite{2014-SubramanyanArora, 2016a-CabodiCamurati.etal,
  2017-CabodiCamurati.etal} by adopting the idea of \emph{taint
  analysis} which originally comes from the software domain. %
This is the research which is the most closely related to our work. %
In those approaches a HW security requirement such as a specific
confidentiality requirement is formulated in terms of a \emph{taint
  property}~\cite{2014-SubramanyanArora} along a certain path in the
design. %
The property fails if the taint can propagate, i.e., the information
can flow from the source of the path to its destination, while
conforming to a propagation condition. %
This is also related to the notion of observational determinism
defined in~\cite{2010-ClarksonSchneider} in the sense that a taint
property checks whether the observations at the end of the taint path
(destination) are functionally independent of values at the source of
the path. %
This was further developed in~\cite{2016a-CabodiCamurati.etal} and a
portfolio of different taint properties in the CTL language was
created. %
In order to formulate the properties in CTL, certain assumptions about
the attack are required which significantly restrict the coverage of
the method. %

As an alternative, a miter-based equivalence checking technique, with some
resemblance to our computational model in
Fig.~\ref{fig:computational-model}, has been used in previous
approaches~\cite{2017-CabodiCamurati.etal,2017-HuArdeshiricham.etal}. %
Although this increases the generality of the proof, it still
restricts the attack to a certain path. %
Moreover, since some of this work considers verification at the
architectural level, vulnerabilities based on microarchitectural side
channels are not detectable. %

Taint analysis by these approaches has shown promise for formal
verification of certain problems in HW security, for example,
for proving key secrecy in an SoC. %
It also proved useful for security analysis in abstract system
models. %
Note, however, that all these approaches mitigate the complexity of
formal security analysis by decomposing the security requirements into
properties along selected paths. %
This demands making assumptions about what paths are suspicious and
requires some ``clever thinking'' along the lines of a possible
attacker. %
As a result, non-obvious or unprecedented side channels may be
missed. %
 
Any leakage determined by a taint property will also be found using
UPEC because any counterexample to the taint property is also a
counterexample to the UPEC property of Eq.~\ref{eq:upec-property}. %
The important advantage of UPEC is, however, that it does not need any
specification of the expected leakage path. %

\vspace{+0.7ex}
\textit{Micro-architectural side channel attacks} are a security threat
that %
 cannot be discovered by looking only at software and
verifying hyperproperties or taint properties. %
In order to secure the hardware against side-channel attacks at
design time, the notion of non-interference has been adopted %
 in~\cite{2013-WasselGao.etal,2011-TiwariOberg.etal} to prove
the absence of side channels in a hardware design. %
Non-interference in hardware is a strong and also conservative
security requirement. %
However, %
the amount of incurred overhead is
usually prohibitive in many hardware design scenarios. %
Designers usually try to quantitatively evaluate the existence and
leakiness of side channels to %
make a %
 leak/risk
tradeoff~\cite{2014-DemmeSethumadhavan}. %
Such quantitative approaches usually either use simulation-based
methods to collect and analyze a set of execution
traces~\cite{2012-DemmeMartin.etal} or estimate leaks through
mathematical modeling~\cite{2010-DomnitserAbughazel.etal}. %

The whole picture has been significantly altered after the emergence
of Meltdown and Spectre, which proved the existence of a new class of
side-channel attacks that do not rely on a valid execution of a victim
software and can be carried out even if there is no vulnerability in
the deployed software. %
The new findings prove the inadequacy of existing verification
techniques and call for new methods capable of addressing the new
issues. %
While the software verification methods are not capable of finding
covert channel attacks %
because they abstract away the %
 whole
microarchitecture, hardware taint property techniques %
came %
 also short
of capturing these vulnerabilities due to path limitation, as mentioned
above. %

This paper addresses the issue of covert channel attack
vulnerabilities by proving unique program execution.  %
UPEC defines the
requirement for security against covert channel attacks in
microarchitecture designs at the RTL (in which most
of the related vulnerabilities appear) and provides a methodology to
make proofs feasible for mid-size processors. %
Unlike the above techniques for software verification which try to verify a given software,
UPEC models software symbolically so that it exhaustively searches for
a program exposing a HW security vulnerability, or proves that
no such program exists. %

Language-based security is another line of research which advocates the use of
more expressive security-driven hardware description languages. %
  SecVerilog~\cite{2015-ZhangWang.etal} extends the Verilog language
  with a security type system. %
  The designer needs to label storage elements with security types,
  which enables enforcing information flow properties. %
Although using Verilog as the base of the language eases the adoption
of the method, %
the labeling process is complicated %
 and
the designer may need to relabel the design %
in order %
to verify different
security properties. %
All in all, system-wide labeling of an RTL design with thousands of
state bits usually is not a trivial task. %

Program synthesis techniques have been proposed to target HW
security vulnerabilities related to Meltdown, Spectre and their
variants~\cite{2018-TrippelLustig.etal}. %
With the help of these techniques, program tests capable of producing
a specific execution pattern %
for a known attack %
can be generated automatically. %
In order to evaluate security for a certain microarchitecture against
the attack, the user needs to provide a formal description of the
microarchitecture and the execution pattern associated with the
attack. %
The synthesis tool can then synthesize the attacker program which can
be later used to test the security of the computing system against the
considered attack. %
Although the method automates the generation of an attacker program,
the execution pattern of the considered attack must be specified by
the user, i.e., the user still needs to develop the attack in an
abstract way by reasoning along the lines of a possible attacker. %
Furthermore, the coverage of the generated tests are restricted to
known attacks and vulnerabilities. %

\section{Orc: A New Kind of Covert Channel Attack}
\label{sec:raw-hazard-attack}

The term \emph{Read-After-Write (RAW) Hazard} denotes a well-known
design complication resulting directly from pipelining certain
operations in digital hardware~\cite{2013-PattersonHennessy}. %
The hazard occurs when some resource is to be read after it is
written causing a possible stall in the pipeline. %
RAW hazards exist not only in processor execution pipelines but also
elsewhere, e.g., in the core-to-cache interface. %

For reasons of performance, many cache designs employ a pipelined
structure which allows the cache to receive new requests while still
processing previous ones. %
This is particularly beneficial for the case of store instructions,
since the core does not need to wait until the cache write transaction
has completed. %
However, this can create a RAW hazard in the cache pipeline, if a load
instruction tries to access an address for which there is a pending
write. %

A RAW hazard needs to be properly handled in order to ensure that the
correct values are read. %
A straightforward implementation uses a dedicated hardware unit called
\emph{hazard detection} that checks for every read request whether or
not there is a pending write request to the same cache line. %
If so, all read requests are removed until the pending write has
completed. %
The processor pipeline is stalled, repeating to send read requests
until the cache interface accepts them. %

In the following, we show an example how such a cache structure can
create a security vulnerability allowing an attacker to open a covert
channel. %
Let's assume we have a computing system with a cache with
write-back/write-allocate policy and the RAW hazard resolution just
described. %
In the system, some confidential data (\textit{secret data}) is stored
in a certain protected location (\textit{protected address}). %

For better understanding of the example, let us make some more
simplifying assumptions that, however, do not compromise the
generality of the described attack mechanism. %
We assume that the cache holds a valid copy of the secret data (from
an earlier execution of privileged code). %
We also simplify by assuming that each cache line holds a single byte,
and that a cache line is selected based on the lower 8 bits of the
address of the cached location. %
Hence, in our example, there are a total of $2^{8}=256$ cache lines. %

\begin{figure}
  \begin{center}
  \ttfamily
  \scriptsize
  \begin{tabbing}
    XXXXX\=XXXXXXXXXXXXXXXXXXXXXXXXXX\=\kill
    \>  1:\' li x1, \#protected\_addr \>// x1 $\leftarrow$ \#protected\_addr \\
    \>  2:\' li x2, \#accessible\_addr \>// x2 $\leftarrow$ \#accessible\_addr \\
    \>  3:\' addi x2, x2, \#test\_value \>// x2 $\leftarrow$ x2 + \#test\_value\\
    \>  4:\' sw x3, 0(x2) \>// mem[x2+0] $\leftarrow$ x3 \\
    \>  5:\' lw x4, 0(x1) \>// x4 $\leftarrow$ mem[x1+0]\\
    \>  6:\' lw x5, 0(x4) \>// x5 $\leftarrow$ mem[x4+0]\\
  \end{tabbing}
  \caption{Example of an Orc attack: in this RISC-V code
    snippet, \textit{accessible\_addr} is an address within the
    accessible range of the attacker process. %
    Its lower 8 bits are zero. %
The address within its 256 bytes offset
    is also accessible. %
    \textit{test\_value} is a value in the range of 0\ldots 255.%
}
  \label{fig:raw-attack}
  \vspace{-4ex}
\end{center}
\end{figure}

The basic mechanism for the Orc attack is the following. %
Every address in the computing system's address space is mapped to
some cache line. %
If we use the secret data as an address, then the secret data
also points to some cache line. %
The attacker program ``guesses'' which cache line the secret data
points to. %
It sets the conditions for a RAW hazard in the pipelined cache
interface by writing to the guessed cache line. %
If the guess was correct then the RAW hazard occurs, leading to
slightly longer execution time of the instruction sequence than if the
guess was not correct. %
Instead of guessing, of course, the attacker program iteratively tries
all 256 possible cache locations until successful. %

Fig.~\ref{fig:raw-attack} shows the instruction sequence for one such
iteration. %
The shown \ASSCODE{\#test\_value} represents the current guess of the
attacker and sets the RAW hazard conditions for the guessed cache
line. %
The sequence attempts an illegal memory access in instruction~\#5 by
trying to load the secret data from the protected address into
register~\REGX{4}. %
The processor correctly intercepts this attempt and raises an
exception. %
Neither is the secret data loaded into~\REGX{4} nor is instruction~\#6
executed because the exception transfers control to the operating
system with the architectural state of instruction~\#5. %
However, before control is actually transferred, instruction~\#6 has
already entered the pipeline and has initiated a cache transaction. %
The cache transaction has no effect on the architectural state of the
processor. %
But the execution time of the instruction sequence depends on the
state of the cache. %
When probing all values of \ASSCODE{\#test\_value}, the case will
occur where the read request affects the same cache line as the
pending write request, thus creating a RAW hazard and a stall in the
pipeline. %
It is this difference in timing that can be exploited as a side
channel. %

Let us look at the sequence in more detail. %
The first three instructions are all legal for the user process of the
attacker. %
Instruction~\#1 makes register~\REGX{1} a pointer to the secret
data. %
Instruction~\#2 makes register~\REGX{2} a pointer to some array in the
user space. %
The address of this array is aligned such that the 8 least significant
bits are 0. %
Instruction~\#3 adds \ASSCODE{\#test\_value} as an offset
to~\REGX{2}. %
This value is in the range 0\ldots 255. %

Instruction~\#4 is a \emph{store} of some value \REGX{3} into the user
array at~\REGX{2}. %
This is a legal instruction that results in a write request to the
cache. %
Note that the destination cache line is determined by
\ASSCODE{\#test\_value} since the lower 8 bits of the write address
are used for cache line addressing
(cf.~Sec.~\ref{sec:raw-hazard-attack}). %
The cache accepts the write request and immediately becomes ready for
a new request. 
The cache controller marks this write transaction as pending until it
is complete. %
(This takes a few clock cycles in case of a cache hit and even
significantly longer in case of a miss.)
  
In the next clock cycle, instruction~\#5 attempts an illegal load from
the secret address, producing a read request to the cache. %
It will take a small number of clock cycles before the instruction has
progressed to the write-back (WB) stage of the processor pipeline. %
In this stage the exception will be raised and control will be
transferred to the OS kernel. %
Until the instruction reaches the WB stage, however, all components
including the cache keep working. %
Since the cache has a valid copy of the secret data, it instantly
answers the read request and returns the secret data to the core where
it is stored in some internal buffer (inaccessible to software). %
  
Even though the illegal memory access exception is about to be raised,
instruction~\#6 is already in the pipeline and creates a third read
request to the cache. %
This request is created by a forwarding unit using the value stored in
the internal buffer. %
Note that the request is for an address value equal to the secret
data. %
(It does not matter whether an attempt to load from this address is
legal or not.)

Let us assume that this address value happens to be mapped to the same
cache line as the write request to the user array from
instruction~\#4. %
This will be a read-after-write (RAW) hazard situation, in case the
write request from instruction~\#4 is still pending. %
The read transaction must wait for the write transaction to finish. %
The cache controller stalls the core until the pending write
transaction has completed. %
  
In case that the read request affects a different cache line than the
pending write request there is no RAW hazard and the processor core is
not stalled. %

In both cases, the processor will eventually raise the exception and
secret data will not appear in any of the program-visible registers. %
However, the execution time of the instruction sequence differs in the
two cases because of the different number of stall cycles. %
The execution time depends on whether or not a RAW hazard occurs,
i.e., whether or not \ASSCODE{\#test\_value} is equal to the 8 lower
bits of the secret data.%

Assuming the attacker knows how many clock cycles it takes for the
kernel to handle the exception and to yield the control back to the
parent process, the attacker can measure the difference in execution
time and determine whether the lower 8 bits of the secret are equal to
\ASSCODE{\#test\_value} or not. %
By repeating the sequence for up to 256 times (in the worst case), the
attacker can determine the lower 8 bits of the secret. %
If the location of the secret data is byte-accessible, the attacker
can reveal the complete secret by repeating the attack for each byte
of the secret. %
Hardware performance counters can further ease the attack since they
make it possible to explicitly count the number of stalls. %

This new covert channel can be illustrated at the example of the
\RISCV{} RocketChip~\cite{2016-AsanovicAvizienis.etal}. %
The original RocketChip design is not vulnerable to the Orc
attack. %
However, with only a slight modification (17 lines of code (LoC) in an
RTL design of $\sim$250,000~LoC) and without corrupting the
functionality, it was possible to insert the vulnerability. %
The modifications actually optimized the performance of the design by
bypassing a buffer in the cache, by which an additional stall between
consecutive load instructions with data dependency was removed. %
There was no need to introduce any new state bits or to change the
interface between core and cache. %
  The attack does not require the processor to start from a specific state: any program can precede the code snippet of Fig.~\ref{fig:raw-attack}. The only requirement is that \texttt{\small protected\_addr} and \texttt{\small accessible\_addr} reside in the cache before executing the code in Fig.~\ref{fig:raw-attack}. %

The described vulnerability is a very subtle one, compared to Meltdown
and Spectre. %
It is caused by a RAW hazard not in the processor pipeline itself but
in its interface to the cache. %
It is very hard for a designer to anticipate an attack scenario based
on this hazard. %
The timing differences between the scenarios where the RAW hazard is
effective and those where it isn't are small. %
Nevertheless, they are measurable and can be used to open a covert
channel. %

This new type of covert channel discovered by UPEC gives some
important messages: 
 
\begin{itemize}
\item Subtle design changes in standard RTL processor designs, such as
  adding or removing a buffer, can open or close a covert channel. %
  This raises the question whether Spectre and Meltdown are only the
  tip of the iceberg of covert channels existing in today's designs. %
  Although specific to a particular design, the newly discovered vulnerabilities may
  inflict serious damage, once such a covert channel becomes known in
  a specific product. %
\item %
  The \emph{Orc} attack is based on the interface between the core (a
  simple in-order core in this case) and the cache. %
  This provides the insight that the
  \textit{\underline{orc}hestration} of component communication in an
  SoC, such as RAW hazard handling in the core-to-cache interface, may
  also open or close covert/side channels. %
  Considering the complex details of interface protocols and their
  implementation in modern SoCs, this can further complicate verifying
  security of the design against covert channel attacks.
\item The new insight that the existence of covert channels does not
  rely on certain types of processors but on decisions in the RTL
  design phase underlines the challenge in capturing such
  vulnerabilities and calls for methods dealing with the high
  complexity of RTL models. %
\item The presented attack is based on a so far unsuspicious
  microarchitectural feature as its covert channel. %
  This makes it resistant to most existing techniques of security
  verification, as discussed in Sec.~\ref{sec:related-work}. %
  The verification method, therefore, should be exhaustive and must
  not rely on \textit{a~priori} knowledge about the possible
  attacks. %
\end{itemize}

These challenges motivate the proposed UPEC approach. %
It is meant to be used by the designer during the RTL design phase to
detect all possible cases of a covert channel. %

\section{Unique Program Execution Checking (UPEC)}
\label{sec:upec}

\emph{Confidentiality} in HW/SW systems requires that an
untrusted user process must not be able to read %
 protected secret
data. %
In case of a microarchitectural covert channel attack, the attacker
cannot read the secret data directly. %
Nevertheless, confidentiality is violated because the execution timing
of the attacker process depends on the secret data, and the timing
information is measurable, e.g., through user-accessible counters. %
These timing differences may stem from various sources that need to be
exhaustively evaluated when verifying confidentiality. %

In the following, we refer to the computing system to be analyzed for
security as System-on-Chip (SoC) and divide its state variables into
two sets: state variables associated with the content of its memory
(main memory and memory-mapped periphery) and state variables
associated with all other parts of the hardware, the \emph{logic
  parts}. %

  \begin{definition}[Microarchitectural State Variables]
    \label{def:microarchitectural-state-variables}
    \mbox{}\\
    The \emph{microarchitectural state variables} of an SoC are the
    set of all state variables (registers, buffers, flip-flops)
    belonging to the \emph{logic part} of the computing system's
    microarchitecture. %
    \hfill$\Box$%
  \end{definition}

  A subset of these microarchitectural state variables are
  \emph{program-visible}: %

  \begin{definition}[Architectural State Variables]
    \label{def:architectural-state-variables}
    \mbox{}\\
    The \emph{architectural state variables} of an SoC are the subset
    of microarchitectural state variables that define the state of
    program execution at the ISA level (excluding the program state
    that is represented in the program's memory). %
    \hfill$\Box$%
  \end{definition}

\begin{definition}[Secret Data, Protected Location]
    A set of \emph{secret data}~$D$ is the content of memory at a
    \emph{protected location~$A$}, i.e., there exists a protection
    mechanism such that a user-level program cannot access~$A$ to read
    or write~$D$. %
  \hfill$\Box$%
\end{definition}

  The protected location may be in the main memory space, in
  peripherals or in other type of storage in the \emph{non-logic part}
  of the computing system. %
  In addition, there may exist temporary copies of the secret data in
  the cache system. %

\begin{definition}[Unique Program Execution]
  A program \emph{executes uniquely w.r.t. a secret~$D$} if and only
  if the sequence of valuations to the set of architectural state
  variables is independent of the %
values %
 of~$D$, in every clock
  cycle of program execution. %
  \hfill$\Box$ %
\end{definition}

In other words, a user-level program executes uniquely if different
secrets in the protected location do not lead to different values of
the architectural states or to different time points when these values
are assigned. %

\begin{definition}[Confidentiality/Observability]
  A set of secret data~$D$ in a protected location~$A$ is
  \emph{confidential} if and only if any user-level program executes
  uniquely w.r.t.~$D$. %
  Otherwise $D$~is \emph{observable}. %
  \hfill$\Box$%
\end{definition}

Based on this definition, confidentiality is established by proving
unique program execution at the RTL. %
In UPEC this is analyzed by a mathematically rigorous, ``formal'',
method. %
The requirement of unique program execution is formalized as a
``property'' expressed in a property language which is understood by a
(commercially available) tool for property checking. %
Current tools for property checking, however, are built for functional
design verification. %
In order to make property checking applicable to UPEC, we present a
tailor-made computational model and formulate a specific property to
be proven on this model. %

\begin{figure}
  \centering
  \includegraphics[trim={32mm 50mm 90mm 30mm}, clip, width=0.7\linewidth]{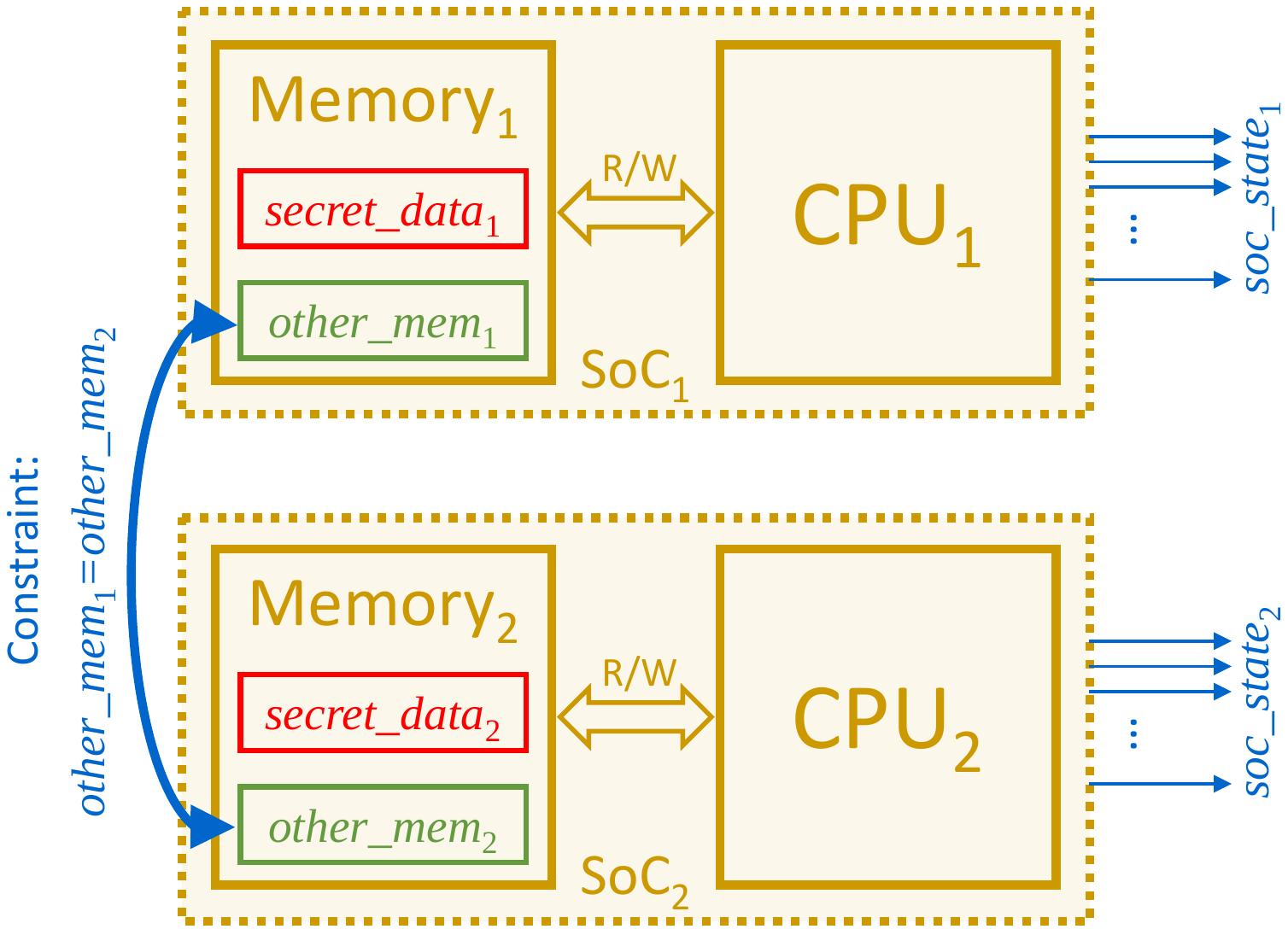} 
  \caption{Computational model for UPEC: two almost identical instances of the computing system are considered. They contain the same data, except for the secret data that may vary between the two instances. The memory block in this figure stands for any resource that may carry secret data including peripheral devices.}
  \label{fig:computational-model}
\end{figure}

Fig.~\ref{fig:computational-model} shows the model that is used in our
UPEC approach. %
It can be derived automatically from the RTL description
of the design and only requires the user to provide the protected
memory region. %
\newcommand{\SOC}[1]{\CODEVARSUB{SoC}{#1}}
\newcommand{\MEMORY}[1]{\CODEVARSUB{Memory}{#1}} %
In this model, \SOC{1} and \SOC{2} are two identical instances of the
logic part of the SoC under verification. %
\MEMORY{1} and \MEMORY{2}, as indicated in the figure, hold the same
set of values except for the memory location of a defined secret
data. %

Based on this model, we propose the \emph{UPEC property}: %
For a system to be secure %
w.r.t. %
 covert channel attacks, the computational model derived from the design's RTL description has to
fulfill the following property expressed in the CTL property language~\cite{1982-EmersonClarke}: 
\newcommand{\SECRETDATAPROTECTED}{\CODEVAR{secret\_data\_protected}}
\newcommand{\SOCSTATE}[1]{\CODEVARSUB{soc\_state}{#1}}
\newcommand{\MICROSOCSTATE}[1]{\CODEVARSUB{micro\_soc\_state}{#1}}
\begin{align}
  \label{eq:upec-property}
  \CTLAG{}\, (\SECRETDATAPROTECTED{}\nonumber\\
  \wedge\, \MICROSOCSTATE{1} &= \MICROSOCSTATE{2}\nonumber\\
  \to\, \CTLAG{}\, \SOCSTATE{1}&=\SOCSTATE{2})
\end{align}

In this formulation, \MICROSOCSTATE{} is a vector of all
  microarchitectural state variables, as defined in
  Def.~\ref{def:microarchitectural-state-variables}, \SOCSTATE{} is
  some vector of state variables which includes, as a subset,
  all architectural state variables as defined in
Def.~\ref{def:architectural-state-variables} but not
  necessarily all other microarchitectural state variables. %
The constraint \SECRETDATAPROTECTED{} specifies that a protection
mechanism in the hardware is enabled for the secret data memory
location. %
The CTL operators \CTLAG{} denote that the following condition must be
fulfilled at all times and for all possible runs of the system
(``\CTLOPERATOR{A}ll paths, \CTLOPERATOR{G}lobally''). %

The property in Eq.~\ref{eq:upec-property} fails if and only if, in
the system under verification, there exists a state, \SOCSTATE{}, such
that the transition to the next state, \SOCSTATE{}', depends on the
secret data. %
This covers all situations in which program execution is not unique. %
Commercial property checking tools are available to check such a
property based on standardized languages like SystemVerilog Assertions
(SVA)~\cite{2009-SystemVerilog}. %
For reasons of computational complexity, however, standard solutions
will fail so that a method specialized to this problem has been
developed, as described in Sec.~\ref{sec:upec-on-a-bounded-model}. %

Importantly, in our methodology we will consider situations where
\SOCSTATE{}, besides the architectural state variables of the SoC, includes some or all microarchitectural state variables, such
as the pipeline buffers. %
Producing a unique sequence for a superset of the architectural state
variables represents a sufficient but not a necessary condition for
unique program execution. %
This is because secret data may flow %
to microarchitectural registers which are not observable by the user
program, i.e., they do not change program execution at any time, and,
hence, no secret data is \emph{leaked}. %

We therefore distinguish the following kinds of counterexamples to the
UPEC property:

\begin{definition}[\LALERT{}]
  \label{def:l-alert}
  \mbox{}\\
  A \emph{leakage alert (\LALERT{})} is a counterexample leading to a
  state with $\SOCSTATE{1} \neq \SOCSTATE{2}$ where the differing
  state bits are \emph{architectural} state variables. %
  \hfill$\Box$%
\end{definition}

\LALERTS{} indicate that secret data can affect the sequence of
architectural states. %
This reveals a direct propagation of secret data into an architectural
register (that would be considered a \emph{functional} design bug),
or a more subtle case of changing the timing and/or the values of the
sequence without violating the functional design correctness and
without leaking the secret directly. %
UPEC will detect the HW vulnerability in both cases. %
While the former case can be covered also by standard methods of
functionally verifying security requirements, this is not possible in
the latter case. %
Here, the opportunity for a covert channel attack may be created, as
is elaborated for the Orc attack in
Sec.~\ref{sec:raw-hazard-attack}. %
Since the functional correctness of a design is not violated, such a
HW vulnerability will escape all traditional verification methods
conducted during the microarchitectural design phase.%

\begin{definition}[\PALERT{}]
  A \emph{propagation alert (\PALERT{})} is a counterexample leading to
  a state with $\SOCSTATE{1} \neq \SOCSTATE{2}$ where the differing
  state bits are microarchitectural state variables %
that are not %
 architectural state variables. %
  \hfill$\Box$ %
\end{definition}

\PALERT{}s show possible propagation paths of secret data from the
cache or memory to program-invisible, internal state variables of the
system. %
A \PALERT{} very often is a precursor to an \LALERT{}, because the
secret often traverses internal, program-invisible buffers in the
design before it is propagated to an architectural state variable like
a register in the register file. %

The reason why \SOCSTATE{} in our methodology may also include
program-invisible state variables will be further elaborated in the
following sections. %
In principle, our method could be restricted to architectural state
variables and \LALERTS{}. %
\PALERTS{}, however, can be used in our proof method as early
indicators for a possible creation of a covert channel. %
This contributes to mitigating the computational complexity when
proving the UPEC property. %

\section{UPEC on a bounded model}
\label{sec:upec-on-a-bounded-model}

Proving the property of Eq.~\ref{eq:upec-property} by classical
unbounded CTL model checking is usually infeasible for SoCs of
realistic size. %
Therefore, we pursue a SAT-based approach based on ``\emph{any-state
  proofs}'' in a bounded circuit model. %
This variant of Bounded Model Checking
(BMC)~\cite{1999-BiereCimatti.etal} is called Interval Property
Checking (IPC)~\cite{2008-NguyenThalmaier.etal} and is applied to the
UPEC problem in a similar way as in~\cite{2018-FadihehUrdahl.etal} for
functional processor verification. %

For a full proof of the property in Eq.~\ref{eq:upec-property} by our
bounded approach, in principle, we need to consider a time window as
large as the sequential depth, \DSOC{}, of the logic part of the
examined SoC. %
This will be infeasible in most cases. %
However, employing a symbolic initial state enables the solver to
often capture hard-to-detect vulnerabilities within much smaller time
windows. %
A violation of the UPEC property is actually guaranteed to be
indicated by a \PALERT{} in only a single clock cycle needed to
propagate secret data into some microarchitectural state variable of
the logic part of the SoC. %
In practice, however, it is advisable to choose a time window for the
bounded model which is as long as the length, \DMEM{}, of the longest
memory transaction. %
When the secret is in the cache, \DMEM{} is usually the number of
clock cycles for a cache read. %
When the secret is not in the cache, \DMEM{} is the number of clock
cycles the memory response takes until the data has entered the cache
system, e.g., in a read buffer. %
This produces \PALERTS{} of higher diagnostic quality and provides a
stronger basis for inductive proofs that may be conducted
subsequently, as discussed below.%

\subsection{Dealing with spurious counterexamples}
\label{sec:spurious-counterexamples}

In the IPC approach, the symbolic initial state also includes
unreachable states of the system. %
This may result in counterexamples which cannot happen in the normal
operation of the system after reset (\emph{spurious
  counterexamples}). %
This problem is usually addressed by strengthening the symbolic
initial state by \emph{invariants}, in order to exclude certain
unreachable states. %
However, developing invariants is mainly done in an ad-hoc way and it
is not a trivial task to find sufficient invariants in a large SoC. %
However, the UPEC computational model and property formulation make it
possible to address the problem of spurious counterexamples in a
structured way. %
Since both SoC instances start with the same initial state, all of the
unreachable initial states and spurious behaviors have the same
manifestation in both SoC instances and therefore do not violate the
uniqueness property, except for the states related to the memory
elements holding the secret value. %
Here, we address the problem by adding three additional constraints on
the symbolic initial state. %

\textit{Constraint~1%
  , ``no on-going protected accesses''%
}. %
Protected memory regions are inaccessible to certain user processes
but the OS-level kernel process can freely access these regions. %
This can create a scenario for confidentiality violation in which the
kernel process loads secret data from a protected region to a
general-purpose register and then instantly branches into a malicious
user-level process. %
Such an explicit revelation of secret data can be considered as
spurious counterexample and excluded from consideration since it
cannot happen in a SoC running an operating system within its normal
operation. %
To exclude these trivial leakage scenarios, the proof must be
constrained to exclude such %
initial states that implicitly represent ongoing memory transactions
in which protected memory regions are accessed. %
This constraint can be formally expressed by assuming that, at the
starting time point, the buffers holding the addresses for ongoing
memory transactions do not contain protected addresses. %
These buffers can be easily identified by inspecting the fan-in of the
memory interface address port. %

\textit{Constraint 2%
  , ``cache I/O is valid''%
}. %
Since the cache can hold a copy of the secret data, an unreachable
initial state for the cache controller may lead to a spurious
counterexample where the secret data is leaked in an unprivileged
access. %
Deriving invariants for a complex cache controller in order to exclude
such spurious behavior requires in-depth knowledge about the design
and, also, expertise in formal verification. %
However, the task can be significantly simplified if we assume that
the cache is implemented correctly. %
This is justifiable because UPEC-based security verification should be
carried out on top of conventional functional verification, in order
to target vulnerabilities not corrupting functionality but
compromising security. %

Communication between the cache and the processor and also higher
levels of memory typically happens based on a well-defined protocol. %
Spurious behaviors reveal themselves by violating this protocol. %
Protocol compliance is usually verified as a part of functional
verification (through simulation or formal techniques) and the
requirements for such compliance is usually defined by the protocol
specification and does not require in-depth knowledge about the
implementation. %
For UPEC, we ensure the protocol compliance and valid I/O behavior of
the cache by instrumenting the RTL with a special %
cache monitor %
which observes the transactions between the cache and
other parts of the system (processor, main memory, etc.) and raises a
flag in case of an invalid I/O behavior. %
Developing hardware monitors is standard practice in verification and
does not %
impose %
significant overhead in terms of the effort involved. %

 \textit{Constraint 3, ``secure system software''.} %
  Memory protection mechanisms are meant to protect the content of
  memory from user software, however, high-privilege softwares can
  freely access the contents of memory (including the secret data). %
  Since it is not possible to have the complete OS/kernel in our
  bounded model (due to complexity), this may lead to trivial false
  counterexamples in which the system software copies the secret data
  into an unprotected memory location or architectural state. %
  In order to exclude such trivial cases, a constraint %
is needed restricting the search to systems with secure system software. %
  The system software is considered secure iff, under any possible
  input, there %
is %
 no load instruction accessing secret data, i.e.,
  before any load instruction, there is %
always %
an appropriate bounds check. %
  In order to reflect that in the proof, %
the added constraint specifies that when %
 the processor is in kernel mode, at the ISA level of program execution, there is no load instruction of the system software that accesses the secret. At the microarchitectural level this means that either no load instruction of the system software is executed that accesses the secret, or the execution of such a load instruction is invalid, i.e., the load instruction has been speculatively executed for a mispredicted bound check and is invalid at the ISA level. %

It should be noted that these constraints
do not restrict the generality of our proof. %
They are, actually, invariants of the global system. %
Their validity follows from the functional correctness of the OS and
the SoC. %

\subsection{Mitigating complexity}
\label{sec:mitigating-complexity}

As discussed in Sec.~\ref{sec:upec}, in our computational model
the content of the memory is excluded from the \SOCSTATE{} in the UPEC
property (Eq.~\ref{eq:upec-property}). %
Therefore, the content of the memory influences neither the assumption
nor the commitment of the property and, thus, can be disregarded by
the proof method. %

This observation helps us to mitigate the computational proof
complexity by black-boxing (i.e., abstracting away) the data fields in
the cache tables. %
This significantly reduces the state space of the model while it does
not affect the validity of the proof. %
It is important to note that the \textit{logic parts} of the cache,
such as tag array, valid bits, cache controller, memory interface,
etc., must not be black-boxed since they are part of the
microarchitectural state and they play a key role for the security of
the system. %

In order to ensure that the partly black-boxed cache also conforms
with the assumption made about the memory by the UPEC computational
model, another constraint needs to be added to the property: %

\textit{Constraint~4%
  , ``equality of non-protected memory''%
}. %
For any read request to the cache, the same value must be returned by
the cache in both instances of the SoC, unless the access is made to
the memory location holding the secret data. %
This constraint is always valid for a cache that is functionally
correct and therefore does not restrict the generality of the proof. %

\newcommand{\NOONGOINGPROTECTEDACCESS}{\CODEVAR{no\_ongoing\_protected\_access}}

\subsection{UPEC interval property}
\label{sec:upec-interval-property}

The interval property for UPEC is shown in
Fig.~\ref{fig:ipc-property}. %
The macro \SECRETDATAPROTECTED{}() denotes that in both SoC instances,
a memory protection scheme is enabled in the hardware for the memory
region holding the secret data. %
The macro \NOONGOINGPROTECTEDACCESS{}() defines that the buffers in
each SoC instance holding the addresses for ongoing transactions do
not point to protected locations (Constraint 1). %
\newcommand{\CACHEMONITORVALIDIO}{\CODEVAR{cache\_monitor\_valid\_IO}}
The macro \CACHEMONITORVALIDIO{}() specifies that in each clock cycle during the time
window of the property, the cache I/O behaviors in the two SoC
instances comply with Constraints 2 and~4. %
The last assumption implements Constraint~3.

\begin{figure}
  \centering
  \begin{minipage}{0.9\linewidth}
    \small
    \begin{tabbing}
      XXX\=XXXXXXXXXXX\=XXX\=XXX\=\kill
      \textcolor{blue}{assume:} \\
      \> \textcolor{blue}{at} $t$: \>\SECRETDATAPROTECTED{}(); \\
      \> \textcolor{blue}{at} $t$: \>$\MICROSOCSTATE{1} = \MICROSOCSTATE{2}$; \\
      \> \textcolor{blue}{at} $t$: \> \NOONGOINGPROTECTEDACCESS{}(); \\
      \> \textcolor{blue}{during} $t$..$t+k$: \>\CACHEMONITORVALIDIO{}(); \\ 
      \> \textcolor{blue}{during} $t$..$t+k$: \>\CODEVAR{secure\_system\_software}(); \\ 
      \textcolor{blue}{prove:} \\
      \> \textcolor{blue}{at} $t+k$: \>$\SOCSTATE{1} = \SOCSTATE{2}$; \\
    \end{tabbing}
  \end{minipage}
  \caption{UPEC property (Eq.~\ref{eq:upec-property}) formulated as
    interval property: it can be proven based on a bounded model of
    length $k$ using Satisfiability (SAT) solving and related
    techniques. %
  }
  \label{fig:ipc-property}
\end{figure}

\section{Methodology}
\label{sec:methodology}

\begin{figure}
  \centering
  \includegraphics[trim={19mm 35mm 28mm 5mm}, clip, width=\linewidth]{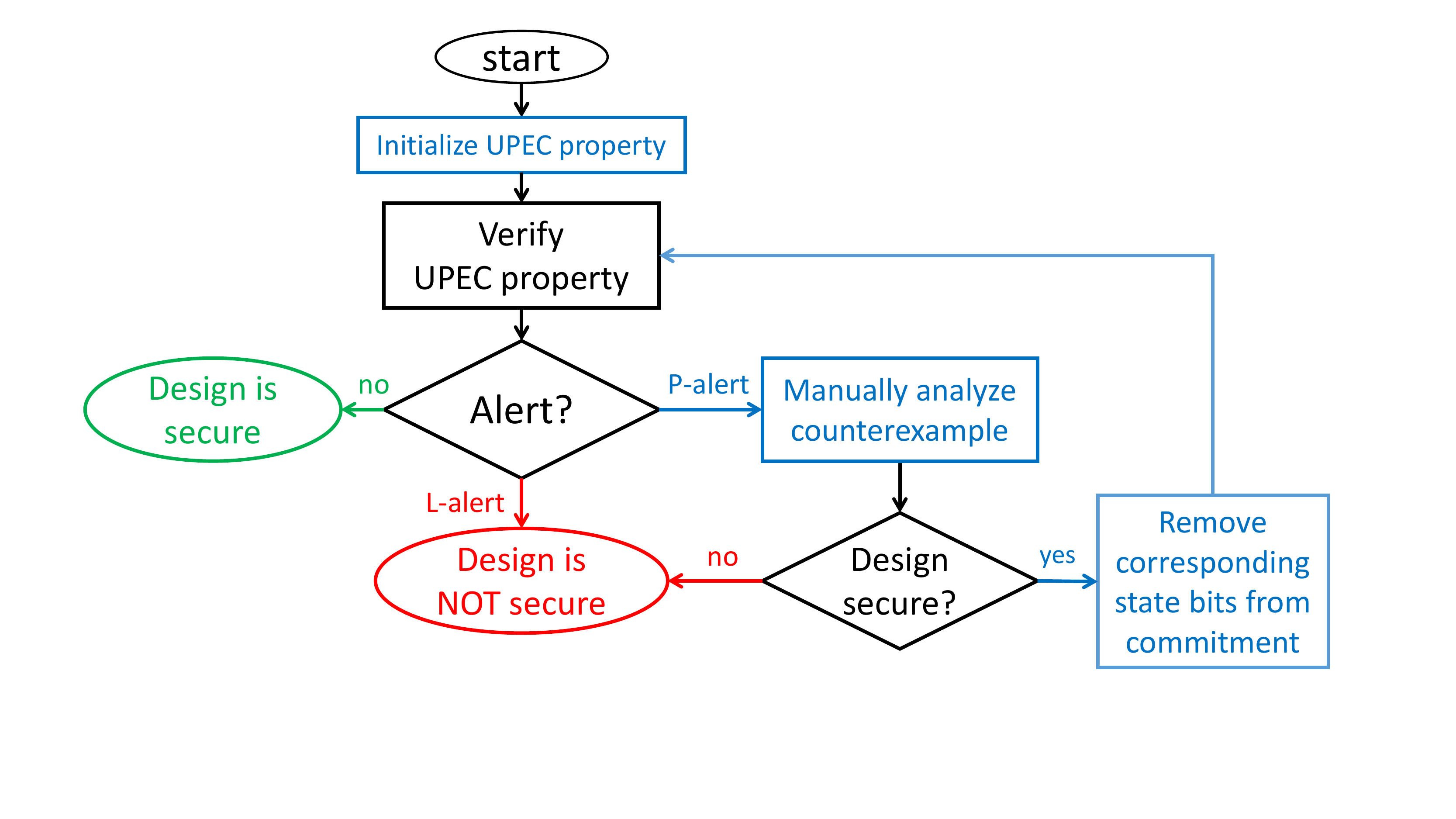} 
  \caption{UPEC Methodology: counterexamples to the UPEC property,
    \PALERTS{} and \LALERTS{}, denote propagation of secret data into
    microarchitectural registers. Only \LALERTS{} prove a security
    violation. \PALERTS{} are only necessary but not sufficient to
    detect a covert channel. However, they can be computed with less
    effort. Based on this iterative process the designer can browse
    through different counterexamples and exploit the trade-off
    between the computational complexity and the diagnostic
    expressiveness of counterexamples to prove or disprove security. }
  \label{fig:upec-flowchart}
\end{figure}

Fig.~\ref{fig:upec-flowchart} shows the general flow of UPEC-based
security analysis of computing systems. %
Checking the UPEC property (Eq.~\ref{eq:upec-property}) is at the core
of a systematic, iterative process by which the designer identifies
and qualifies possible hardware vulnerabilities in the design. %
The UPEC property is initialized on a bounded model of length~\DMEM{}
and with a proof assumption and obligation for the complete set of
microarchitectural state variables. %

If the UPEC property can be successfully verified, then the design is
proven to be free of side effects that can be exploited as covert
channels. %
If the property fails it produces a counterexample which can be either
an \LALERT{} or a \PALERT{}. %
An \LALERT{} exposes a measurable side effect of the secret data on the
architectural state variables, rendering the design insecure. %
A \PALERT{} documents a side effect of the secret data on
microarchitectural state variables that are not directly accessible
by the attacker program. %
In principle, the designer can now remove the affected
microarchitectural state variables from the proof obligation of the
UPEC property (while keeping the complete set of microarchitectural
state variables in the proof assumption), and then re-iterate the process to search for a
different counterexample. %
The process is bound to terminate because, eventually, either an
\LALERT{} occurs or the design is secure. %

In practice, however, the designer will not simply eliminate a
\PALERT{} but instead will analyze the counterexample. %
As mentioned before, an \LALERT{} may have one or several shorter
\PALERTS{} as precursors. %
Since the \PALERT{} is a shorter counterexample than the corresponding
\LALERT{} it can be computed with less computational effort, including
shorter proof times. %
A \PALERT{} belonging to an \LALERT{} documents the earliest
manifestation of a side effect and points the designer to the source
of the vulnerability that causes the side effect. %
If the security compromise is already obvious from the \PALERT{} the
designer may abort the iterative process of
Fig.~\ref{fig:upec-flowchart} %
by deeming the \PALERT{} as ``insecure'' %
and change the RTL in order to remove the vulnerability. %
This may be as simple as adding or removing a buffer. %
  Note that if the designer wrongly deems a \PALERT{} as ``secure'' then
  the security compromise is detected by another \PALERT{} later, or,
  eventually, by an \LALERT{}. %
  The penalty for making such a mistake is the increase in run times
  for checking the later UPEC property instances. %

 If the procedure terminates without producing an \LALERT{} this is not a complete
proof that the design is secure, unless we increment the
length of the model to~\DSOC{}. %
The alternative is to take the \PALERTS{} as starting point for proving
security by an inductive proof of the property in
Eq.~\ref{eq:upec-property} for the special case of an initial state
derived from the considered \PALERT{}. %
A \PALERT{} can be deemed as secure if the designer knows that the 
values in the affected microarchitectural state variables will not propagate under 
the conditions under which the \PALERT{} has occurred. %
In other words, in a secure \PALERT{}, a certain condition holds which implicitly 
represents the fact that the propagation of secret data to
architectural registers will be blocked. %
In order to conduct the inductive proof the designer must identify
these blocking conditions for each \PALERT{}. %
Based on the UPEC computational model
(Fig.~\ref{fig:computational-model}) the inductive proof checks
whether or not the blocking condition always holds for the system once
the corresponding \PALERT{} has been reached. %

Finally, there is always the conservative choice of making design
changes until no \PALERTS{} occur anymore, thus, establishing full
security for the modified design w.r.t. covert channels.

\section{Experiments}
\label{sec:experiments}

Our new insights into covert channels were obtained by implementing
UPEC in a prototype verification environment for security analysis. %
It interfaces with a commercially available property checking tool,
OneSpin~360~DV-Verify, which is used to conduct the IPC proofs
underlying UPEC. %

We explored the effectiveness of UPEC by targeting different design
variants of RocketChip~\cite{2016-AsanovicAvizienis.etal}, an
open-source \RISCV{}~\cite{2014-WatermanLee.etal} SoC generator. %
The considered RocketChip design is a single-core SoC with an \emph{in-order}
pipelined processor and separate data and instruction level-1
caches. %

All results were obtained using the commercial property checker 
OneSpin~360~DV-Verify on an Intel Core~i7-6700 CPU with 32\,GB of RAM,
running at 3.4\,GHz. %

In order to evaluate the effectiveness of UPEC for capturing
vulnerabilities we targeted the original design of RocketChip and two design variants made
vulnerable to (a) a Meltdown-style attack and (b) an Orc
attack, with only minimal design modifications. %
Functional correctness was not affected and 
the modified designs successfully passed all tests provided by the
\RISCV{} framework. %
UPEC successfully captured all vulnerabilities. %
In addition, UPEC found an ISA incompliance in the \emph{Physical Memory Protection} (PMP) 
unit of the original design. %

For the \textit{Meltdown-style attack} we modified the design such that a cache line
refill is not canceled in case of an invalid access. %
While the %
illegal %
access itself is %
not successful but raises an exception, the cache content is modified and can be analyzed by an
attacker. We call this a Meltdown-style attack since the instruction sequence for 
carrying out the attack is similar to the one reported by~\cite{2018-LippSchwarz.etal}. Note, however,
that in contrast to previous reports we create the covert channel based on an in-order pipeline.  %

For the \textit{Orc attack}, we conditionally bypassed one buffer, as
described in Sec.~\ref{sec:raw-hazard-attack}, thereby creating a
vulnerability that allows an attacker to open a covert timing side
channel.%

In the following experiments, the secret data is assumed to be in a protected
location, $A$, in the main memory. %
Protection was implemented using %
  the %
\textit{Physical Memory Protection (PMP)} %
  scheme %
of the \RISCV{} ISA~\cite{2016-WatermanLee.etal} by configuring the
memory region holding the location~$A$ of the secret data as
inaccessible in user mode.

\subsection{Experiments on the original RocketChip design}
\label{sec:experiments-on-the-Original-Rocketchip-Design}

We conducted experiments on the \emph{original} design for %
two cases: %
(1)~$D$~initially resides in the data cache and main memory, %
  and, %
(2)~$D$~initially only resides in the main memory%
  ; %
cf.~the columns labeled ``$D$ in cache'' and ``$D$ not in cache'' in
Tab.~\ref{tab:p-alert}. %

Separating the two cases and applying UPEC to them individually is
beneficial for the overall efficiency of the procedure, because the
solver does not have to consider both cases implicitly within a single
problem instance. %

\begin{table}
  \centering
  \caption{%
    UPEC methodology experiments: %
    {\normalfont For the original RocketChip
    design the two settings are distinguished whether or not there is
    a valid copy of the secret data in the cache. %
    For each case the computational and manual efforts are reported
    when following the UPEC methodology and analyzing possible
    alerts. %
    } %
  }%
  \label{tab:p-alert}
  \renewcommand{\arraystretch}{1.2}
  \begin{tabular}{lcc}
    \hline 
    \hline 
    &
    \rule{0ex}{2.3ex}%
    $D$ cached &
    $D$ not cached \\
    \hline 
    \rule{0ex}{2.3ex}%
    \DMEM{} &
    5 &
    34 \\
    Feasible~$k$ &
    9 &
    34 \\
    \# of \PALERTS{} &
    20 &
    0 \\
    \# of RTL registers causing \PALERTS{} &
    23  &
    N/A \\
    \rule{0ex}{2.3ex}%
    Proof runtime &
    3 hours &
    35 min \\
    Proof memory consumption &
    4 GB &
    8 GB \\
    Inductive proof runtime &
    5 min &
    N/A \\
    \rule{0ex}{2.3ex}%
    Manual effort &
    10 person days &
    5 person hours \\
    \hline 
    \hline 
  \end{tabular}
\end{table}

For the experiment with $D$ not in the cache, UPEC proves that there
exists no \PALERT{}. %
This means that the secret data cannot propagate to any part of the
system and therefore, the user process cannot fetch the secret data
into the cache or access it in any other way. %
As a result, the system is proven to be secure for the case that $D$
is not in the cache. %
Since the property proves already in the first iteration of the UPEC
methodology that there is no \PALERT{}, the verification can be carried
out within few minutes of CPU time and without any manual analysis. %

For the case that $D$ is initially in the cache, we need to apply
the iterative UPEC methodology (Fig.~\ref{fig:upec-flowchart}) in
order to find all possible \PALERTS{}. %
We also tried to capture an \LALERT{} by increasing the length~$k$ of
the time window, until the solver aborted because of complexity. %
The second row in the table shows the maximum~$k$ that was feasible. %
The following rows show the computational effort for this~$k$. %

Each \PALERT{} means that the secret influences certain microarchitectural
registers. %
It needs to be verified whether or not each \PALERT{} can be extended to an
information flow into program-visible architectural registers. %
As elaborated in Sec.~\ref{sec:methodology}, using standard procedures of
commercially available property checking, we can establish
proofs by mathematical \emph{induction}, taking the \PALERTS{} as the
\emph{base case} of the induction. %
The inductive proofs build upon the UPEC computational model and check
whether or not a state sequence exists from any of the known
\PALERTS{} to any other \PALERT{} or \LALERT{}. %
If no such sequence exists, then the system is proven to be secure. %
In this way, we proved security from covert channels also for the case
when the secret is in the cache. %
The manual effort for this is within a few person days and is small
compared to the total design efforts for processor design that usually
are on the order of person years for processors of medium complexity. %
The complexity of an inductive proof for one selected case of the
\PALERTS{} is shown in Table~\ref{tab:p-alert} as an example. %

\subsection{Experiments on the modified RocketChip designs}
\label{sec:experiments-on-the-modified-rocketchip-designs}

\begin{table}
  \centering
  \caption{%
    Detecting vulnerabilities in modified designs: %
    {\normalfont For two different
    attack scenarios the computational effort and window lengths to
    obtain the alerts for disproving security are listed. %
    }%
  } %
  \label{tab:vulnerability-detection}
  \renewcommand{\arraystretch}{1.2}
  \begin{tabular}{lcc}
    \hline 
    \hline 
    \rule{0ex}{2.3ex}%
    Design variant/vulnerability &
    Orc &
    Meltdown-style \\
    \hline 
    \rule{0ex}{2.3ex}%
    Window length for \PALERT{} &
    2 &
    4 \\
    Proof runtime for \PALERT{} &
    1 min &
    1 min \\
    \rule{0ex}{2.3ex}%
    Window length for \LALERT{}  &
    4 &
    9 \\
    Proof runtime for \LALERT{} &
    3 min &
    18 min \\
    \hline 
    \hline 
  \end{tabular}
\end{table}

Table~\ref{tab:vulnerability-detection} shows the proof complexity for
finding the vulnerabilities in the \emph{modified} designs. %
For each case, the UPEC methodology produced 
meaningful \PALERTS{} and \LALERTS{}. %
When incrementing the window length in search for an \LALERT{}, new
\PALERTS{} occurred which were obvious indications of security
violations. None of these violations exploits any branch prediction
of the RocketChip. %
For example, in the Meltdown-style vulnerability, within seconds UPEC produced a
\PALERT{} in which the non-uniqueness manifests itself in the valid
bits and tags of certain cache lines. %
This is a well-known starting point for side channel attacks so that, in practice, no
further examinations would be needed. %
However, if the designer does not have such knowledge the procedure may be
continued without any manual analysis until an \LALERT{} occurs. This took about 18~min
of CPU time. For the design vulnerable to an Orc attack the behavior was similar,
as detailed in Table~\ref{tab:vulnerability-detection}.

\subsection{Violation of memory protection in RocketChip}
\label{sec:violation-of-memory-protection-in-Rocketchip}

UPEC also found a case of ISA incompliance in the implementation of
the \RISCV{} Physical Memory Protection mechanism in RocketChip. %
PMP in the \RISCV{} ISA is managed by having pairs of address and
configuration registers (PMP entry). %
There are 16 registers for 32-bit addresses, each of them associated
with an 8-bit configuration register for storing access attributes. %

There is also a locking mechanism by which the software can lock PMP
entries from being updated, i.e., only a system reboot can change the
contents of any PMP entry. %
According to the ISA specification, if a memory range and the PMP
entry for the end address is locked, the PMP entry of the start
address must be automatically locked, regardless of the contents of
the associated configuration register. %

This mechanism has not been correctly implemented in the RocketChip,
enabling a modification of the start address of a locked memory range
in privileged mode, without requiring a reboot. %
This is clearly a vulnerability, and a bug with respect to the
specification. %
The detection of this security violation in PMP is actually an
example %
  showing %
that the UPEC property of Eq.~\ref{eq:upec-property} also covers
information leakage through a ``main channel'', i.e., in the case
where an attacker may gain direct access to a secret. %
Such cases can be identified by conventional functional design
verification techniques. %
However, in UPEC they are covered without targeting any security
specification and can be identified in the same verification framework
as side channel-based vulnerabilities. %
In the current version of RocketChip, the bug has already been
fixed. %

\section{Conclusion}
\label{sec:conclusion}

This paper has shown that covert channel attacks are not limited to
high-end processors but can affect a larger range of architectures. %
While all previous attacks were found by clever thinking of a human
attacker, this paper presented UPEC, an automated method to
systematically detect all vulnerabilities by covert channels,
including those by covert channels unknown so far. %
Future work will explore measures to improve the scalability of UPEC
to handle larger processors. %
As explained in Sec.~\ref{sec:upec-on-a-bounded-model}, if there is a
security violation, it is guaranteed that this will be indicated
within a single clock cycle by a \PALERT{}. %
Longer unrollings of the UPEC computational model were only chosen to
facilitate the manual process for diagnosis and the setting up of
induction proofs. %
Therefore, we intend to automate the construction of induction proofs,
as needed for the methodology of Sec.~\ref{sec:methodology}. %
This does not only remove the manual efforts for \PALERT{} diagnosis
and for setting up the induction proofs. %
Also, by merit of this automation, the UPEC computational model can be
restricted to only two clock cycles. %
This drastically reduces the computational complexity. %
In addition, a compositional approach to UPEC will be explored. %

\section*{Acknowledgment}
\label{sec:acknowledgment}
We thank Mark D. Hill (U. of Wisconsin), Quinn Jacobson (Achronix
Semiconductor Corp.) and Simha Sethumadhavan (Columbia U.) for their
valuable feedback. %
The reported research was partly supported by BMBF KMU-Innovativ
01IS17083C (Proforma) and by DARPA. %

\ifCLASSOPTIONcaptionsoff
  \newpage
\fi

\bibliographystyle{IEEEtran}
\bibliography{refs3}

\end{document}